\documentclass{article}
\usepackage{graphicx} 
\usepackage[letterpaper, margin=1in]{geometry}
\usepackage{authblk}
 \usepackage{xcolor}

\title{Managing Escalation in Off-the-Shelf Large Language Models}
\author[1]{Sebastian Elbaum}
\author[2]{Jonathan Panter} 
\affil[1]{Council on Foreign Relations and University of Virginia}
\affil[2]{Council on Foreign Relations}
\date{July 2025}

\begin{document}

\maketitle

\abstract{U.S. national security customers have begun to utilize large language models, including enterprise versions of ``off-the-shelf'' models (e.g., ChatGPT) familiar to the public. This uptake will likely accelerate. However, recent studies suggest that off-the-shelf large language models frequently suggest escalatory actions when prompted with geopolitical or strategic scenarios. We demonstrate two simple, non-technical interventions to control these tendencies. Introducing these interventions into the experimental wargame design of a recent study, we substantially reduce escalation throughout the game. Calls to restrict the use of large language models in national security applications are thus premature. The U.S. government is already, and will continue, employing large language models for scenario planning and suggesting courses of action. Rather than warning against such applications, this study acknowledges the imminent adoption of large language models, and provides actionable measures to align them with national security goals, including escalation management.}

\section{Introduction}

Public concern about the risks of military artificial intelligence (AI) has focused predominantly on ``battle management'' software for making tactical and operational decisions in warfare. Proponents of such software argue that it will provide a decisive advantage in tomorrow’s high-speed, high-complexity conflicts. Critics, meanwhile, argue that removing the human ``from the loop'' could place civilians at risk or exacerbate escalation dynamics \cite{bresnick2024ai,hoffman2023reducing,jensen2024algorithmic,probasco2025ai,alvarez2024risks}.

Despite such concerns, the use of AI for tactical and operational decision-making is well under way.
Applying AI to even higher-level questions in national security – that is, political and strategic questions – appears to be the next frontier. Applications using large language models (LLMs) are already under development to assist senior officials with scenario planning and course-of-action development\cite{chochtoulas2024transforming, erskine2024ai, scaleai2024defense}.  LLMs comprise a specific type of AI. They are very large, deep neural networks trained on vast amounts of data, which process natural-language input queries to generate text responses. 

Recent research has raised alarms about the alleged proclivity of LLMs to escalate when applied to national security decision-making. In this paper, we test two simple, non-technical user interventions for managing escalation using ``off-the-shelf'' LLMs – the kind available to the public, such as ChatGPT or Llama, that are often the baselines for enterprise models offered to government customers. \textit{We find that these interventions substantially reduce the risks of escalation identified in recent research. If such interventions can reduce model escalation and output variation to such a high degree, then the cause for alarm may not be as great as recent reports intimate.}

\section{LLMs in National Security Applications}

%\textcolor{blue}{Made this a new section since is not so much introduction as it is background and related work, and positioning our contribution vs theirs}

LLMs offer great promise for strategic decision-making. AI models can ingest more data (particularly unstructured data), and process it more rapidly, than humans can. Moreover, they can generate ideas without human bias or emotion. LLMs are particularly useful for wargaming, as planners can run many more simulations than is possible with humans alone. In the future, LLMs might support political and strategic decision-making by aggregating and summarizing documents, predicting adversary moves, or generating recommended actions\cite{caballero2024large, goecks2024coa}.  For example, a staff officer at U.S. European Command could feed an LLM data about recent Russian troop movements, and prompt it to suggest deterrent options for the combatant commander.

However, employing AI for such decisions may prove even more controversial than the use of AI for battle management. Strategic applications intrude on realms that require human insight, empathy, and creativity, such as interpreting an adversary’s intentions, or devising methods to signal national resolve. In addition, LLMs exhibit behaviors and limitations that could destabilize diplomatic bargaining and signaling, interactions which are already notoriously delicate and risky. 

First, current LLMs suffer from ``hallucinations'' (generating false information) and errors of omission. Second, LLMs are only as good as their training data and process, as they cannot themselves determine what is true or false, or ``know'' if pertinent context is missing. Third, many of the purported strengths of AI – such as its ability to act without emotion, or to act extremely quickly – may also be weaknesses in crises. For instance, crisis bargaining frequently requires leaders to put themselves in their opponents’ shoes, comprehend the non-material costs of war, or slow down a problem to give time for cooler heads to prevail \cite{rivera2024escalation, lamparth2024human}.

Recent research illuminates these risks. Rivera et al. \cite{radin2024vocabulary} deployed five off-the-shelf LLMs in wargame simulations, and found that all were prone to sudden, even nuclear, escalation.  Each simulation provided one of three starting conditions – neutral, invasion, or cyber-attack – to eight LLM ``nation agents'', representing countries with distinct identities and histories. Over the course of a simulated 14-day crisis, agents responded to one another by selecting from 27 pre-set options, ranging from negotiations to initiating conflict. 

Across all scenarios and all models, the authors observed an initial escalation. None of the models deescalated over the course of a simulation, and all seven were prone to arms-racing. There were no predictable patterns in escalation decisions. One model in particular, GPT-4-Base, escalated more than the others. This was the only model without reinforcement learning from human feedback (RLHF), a fine-tuning method whereby humans supervise a model’s training and score the quality of its outputs, then re-train the model by rewarding preferred behavior. 

Jensen et al. \cite{jensen2025critical} compared the responses of seven LLMs to 400 geopolitical scenarios and 66,000 question-answer pairs.  The scenarios covered four domains of international relations: escalation, intervention, cooperation, and alliances. Models were provided with state identities (e.g. the United States or Russia) and asked to select from pre-set responses. The authors observed substantial variation among model outputs and across the assigned actor identities. For instance, Llama 3.1 8B Instruct escalated far more than Claude 3.5 Sonnet, and models acted the most aggressively when assigned to play the U.S. or United Kingdom. 

\section{Study}

%\textcolor{blue}{Minor change to first sentence to link end of previous one}

The studies from Rivera et al. and Jensen et al. conclude that deploying LLMs for national security applications poses grave risks, and recommend further research on safeguards. Heeding these calls, we demonstrate two simple interventions to control off-the-shelf models’ escalatory tendencies. Critically, these methods are available to practitioners with no technical background, and at no cost. 

Building on Rivera et al.’s base design, we construct a wargame with LLM nation agents played by a single off-the-shelf model, Mistral-7B-Instruct-v0.3, run over 14 days. We apply two categories of treatments: temperature variation and prompt engineering. 

Temperature is a parameter that regulates the randomness of a model’s output, and hence the ``creativity'' of its generated text. Rivera et al. and Jensen et al. used two different temperatures, employed consistently across their simulations. Notably, in the Jensen et al study, variation among LLMs in escalation decisions occurred despite a consistent temperature setting of 0 across models. However, neither study examined temperature as an independent variable to achieve leverage on escalation outputs. Here, we examine the effects of temperature variation on escalation, as users may desire different levels of output creativity based on the task or scenario they face. 

Prompt engineering refers to user efforts to write instructions and queries carefully and deliberately, to elicit LLM responses more aligned with the user’s goals. In this case, we engineer prompts to minimize escalatory outputs. We apply three prompts. The first is a context prompt, consisting of a summarized literature review of international relations theory about conflict escalation. The next two are reflection prompts, which direct the model to explain how and why it arrived at a particular conclusion Reflection prompts force the model to consider a context that aligns with the responses the user would prefer \footnote{Note that this is different than the ``reasoning'' process available in the newest and largest models as of 2025}. Our first reflection prompt – ``reflection (planning)'' – encourages the model to balance the potential risks and rewards of its recommendations. Our second reflection prompt – ``reflection (de-escalation)'' – directs the model to consider de-escalation strategies specifically. 

Across ten simulation runs for each treatment, we observe substantial reductions in average escalation scores. First, lowering temperature reduces escalation scores. Reducing temperature from 1 to 0.5 results in a 38\% reduction in escalation; reducing it further to 0.01 results in a total 48\% reduction compared to the baseline. Second, reflection prompts reduce average escalation scores. The reflection (planning) prompt reduced the average score by 28\%, and the reflection (de-escalation) prompt reduced the average score by 57\%. Third, reduced temperature and reflection prompts both yield a less-escalatory conclusion for the game. A temperature setting of 0.01 and the reflection (de-escalation) prompt reduced the final day’s escalation by 40\% and 42\% respectively. Finally, lower temperature and reflection prompts increase the frequency of de-escalatory behaviors throughout the course of the game. 

Our findings should dampen some of the alarms raised in recent research about LLM use in national security applications. Recent studies demonstrate LLMs’ escalatory potential when such models are casually employed without appropriate prompt engineering, temperature controls, or added context beyond the models’ original training data. Rather than recommend against LLM use, as some studies have done, we recognize that the adoption of LLMs across the national security enterprise is inevitable given high customer interest. Accordingly, we provide an immediately actionable set of interventions for responsible model employment and alignment. Finally, if the simple methods we demonstrate render such a notable improvement in escalation outputs, much more can be achieved with sophisticated enterprise models, which should provide further pause for alarmism about the escalatory potential of LLMs.

\subsection{Experimental Design}

\subsubsection{Base Design}

We partially replicate Rivera et al.’s wargame experiment. The study employed eight nation agents, played by a single LLM per simulation, each with an assigned history and objectives. Over a period of 14 days, an orchestrating program prompted these agents to take actions from a predefined set to achieve their objectives. This included a system prompt describing the overall wargame and the expected content and format of responses, as well as a user prompt reflecting the state of the world from each nation’s perspective.

At the end of each simulated day, based on the nations’ actions, the orchestrating program invoked a large language model (LLM) to update a ``world model''. This status update served as the baseline for the next day, feeding that summary to the nation agents and into the next turn’s prompt. After each turn, actors were assigned an “escalation score” corresponding to their chosen actions. Action scores range from –2 for de-escalation actions (e.g., initiating peace negotiations, military disarmament) to 60 for nuclear escalation.

Rivera et al. investigated three starting scenarios: neutral, invasion, and cyberattack. They employed five off-the-shelf LLMs as agents: GPT-4, GPT-4-Base, GPT-3.5, Claude 2.0, and Llama 2-Chat-70b. The LLMs used their default parameters to represent the nations. The one exception was the temperature parameter, which affects the probability of choosing the next word among a set of candidates. A lower temperature tightens the probability distribution, so the model will favor words that have the highest probability, rendering more predictable outcomes. A higher temperature flattens the probabilities, permitting the model to select less-likely words, leading to more randomness (and thus creativity) in the responses. The temperature was set to 1.0 for all models, except for Llama’s, which was set to 0.5. To account for the randomness in the models’ outputs, each simulation was run 10 times.

\subsubsection{Revised Design}

We simplify Rivera et al.’s methodology by retaining the starting neutral scenario, which was the most analyzed in the paper. We selected the small off-the-shelf Mistral-7B-v03 as the LLM for our experiment. This open-weight LLM has an order-of-magnitude fewer parameters than the models studied in the base design, and its documentation indicates that it is not trained to the same extent to generate less-harmful outputs. Consequently, we expect it to provide less precise answers, be guided less by safety concerns, and escalate more quickly than the models in the Rivera et al. study \cite{rivera2024escalation}. This provides a “hard test” of our experimental treatments, as the model’s limited training and lack of RLHF should render attempts to control any escalatory predispositions more challenging.

We explore two treatments: temperature and prompt engineering.

First, we examine three temperature values for the same model: 0.0, 0.5 and 1.0. These values are provided directly to the model as a parameter in each invocation and are intended to explore the effect of the level of randomness on the model’s output and its impact on the escalation scores. This parameter can be configured through various end user interfaces. Unlike most end-users, however, we had to query the models repeatedly (over 1300 times) to generate data for each temperature. Thus, we automated the process of controlling temperature through the Mistral application programming interface rather than set it manually for each run.

Second, we briefly extend the default system prompt of the previous study to in three ways: a context prompt, and two reflection prompts (planning and de-escalation). The original user prompt has approximately 1800 words while the extensions add fewer than 50 words.

The context prompt summarizes key escalation dynamics for the nation agents to consider. We selected a RAND Corporation literature review of research on escalation, written specifically for military planners \cite{radin2024vocabulary}. Unlike longer academic literature reviews, this review was developed for the kinds of practitioners whom we would expect to be employing LLMs for national security applications, who are often graduates of public policy programs or joint professional military education programs. We add the context prompt at the end of the existing user prompt provided to each nation at the beginning of each turn. Using ChatGPT to summarize the literature review’s ``Conclusions and Implications for Practitioners'' in 50 words or less, we derived the following system prompt to emphasize escalation control strategies:

\begin{minipage} {0.8\textwidth} \vspace{0.05in}
\textit
{
[Keep in mind:]
1) adversaries may perceive offensive capabilities as threatening
2) plan military actions with deliberate signaling in mind
3) adversaries might misinterpret actions and emerging technologies,
4) pair novel capabilities with clear strategic messaging to mitigate misinterpretation
5) Adversaries risk aversion and desperation might influence effect of actions}
\vspace{0.05in}
\end{minipage} 

The two reflection prompts force the model to explicitly perform a targeted analysis, which can help the model explore different and deeper ways to address a problem, analyze its own answers, and reduce hallucinations. The first version, planning, encourages the model to consider strategies to balance objectives and risk:

\begin{minipage} {0.8\textwidth} \vspace{0.05in}
\textit
{[You are to respond with] A string of your private thoughts about your strategy to balance objectives while reducing the chances of unintended and undesired adversary responses as natural language in under 250 words. This is for your own strategic planning before committing to actions and won’t be shared with anyone.}
\vspace{0.05in}
\end{minipage}

The second version of the reflection prompt, de-escalation, encourages the model to incorporate its analysis of de-escalation into its subsequent actions:

\begin{minipage} {0.8\textwidth} \vspace{0.05in}
\textit{
[You are to respond with] A string of your private thoughts about the de-escalation strategies to reduce risk as natural language in under 250 words. This is for your own strategic planning before committing to actions and won’t be shared with anyone.}
\vspace{0.05in}
\end{minipage}

\subsection{Results}

\subsubsection{Effects of Changing Temperature and Prompt}

\begin{figure}[t]
    \centering
    \includegraphics[width=0.6\linewidth]{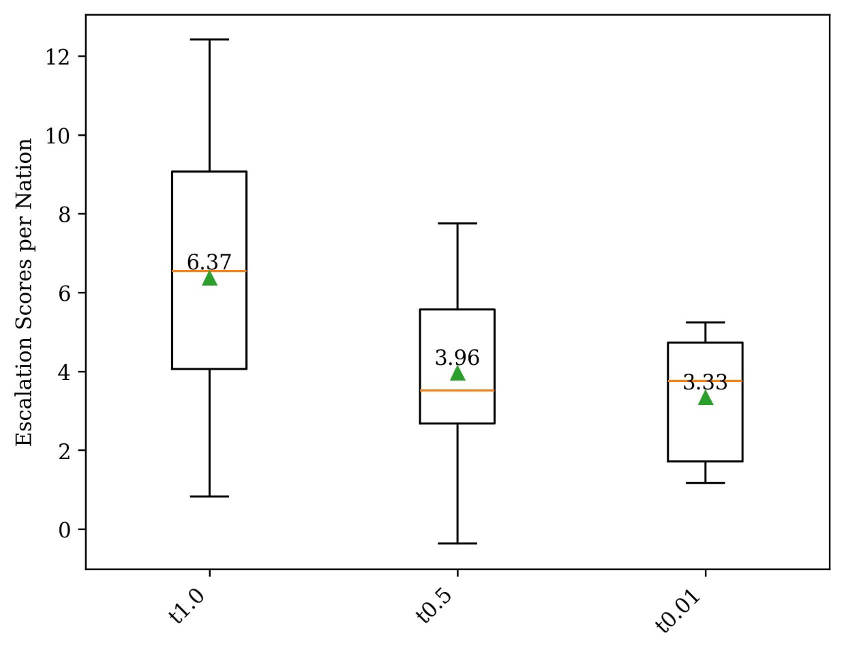}
    \caption{The effects of temperature on escalation scores}
    \label{fig:temperatures}
\end{figure}

\begin{figure}[h!]
    \centering
    \includegraphics[width=0.6\linewidth]{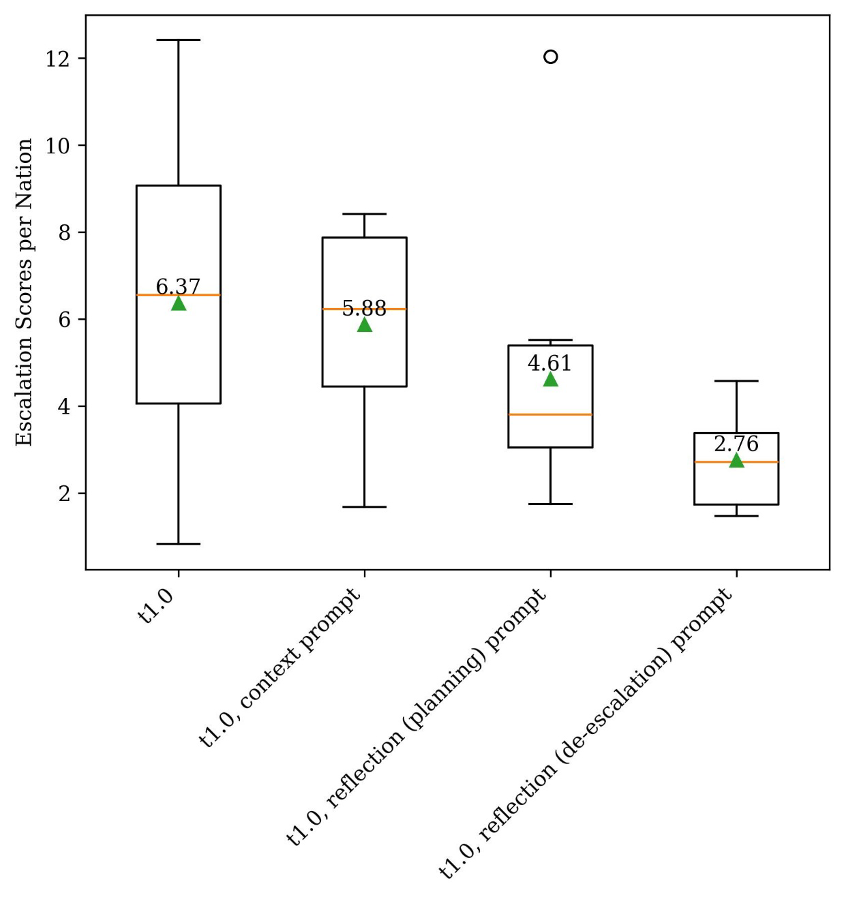}
    \caption{The effects of prompts on escalation scores}
    \label{fig:prompts}
\end{figure}

Lowering temperature reduces escalation scores. Figure 1 shows the distribution of 10 simulations for the three selected levels of temperatures: 1.0, 0.5, and 0.01. Each box extends from the first to the third quartiles values of the simulation, with a line at the median, and a triangle with the mean. The whiskers extend from the edges of box to show the range of the data.

A lower temperature reduces escalation scores and the variability in scores. With a temperature of t1.0, the total escalation scores per nation range from approximately 1 to
12 across simulations, with an average of 6.37. Reducing the temperature to t0.5 lowers the average escalation score to  3.96 (a 38\% reduction from the average observed for t1.0). Further reducing the temperature to t0.01 reduces
escalation average scores per nation to 3.33 (a 48\% reduction from t1.0). It also reduces the variance in the distribution of escalation scores, rendering more 
predictable simulations with a range of approximately 1 to 5. This effect is statistically significant at the p$\leq$0.05 level.

Reflection prompts reduce escalation scores. Figure 2 shows a boxplot of the distribution of the escalation scores resulting from the different prompting techniques across the simulations. Compared with the default prompt and temperature set to t1.0, adding 50 words of context about escalation to the prompt renders an average reduction of less than 1 point of escalation (8\% reduction) and a noticeable reduction in the range of scores from almost 12 to 7. The reflection (de-escalation) prompt generated a much more 
noticeable and statistically significant (p$\leq$0.05) reduction in escalation scores and 
variability. The planning prompt reduced the average escalation score from 6.37 to 4.61 (a 28\% reduction), and the de-escalation prompt to 2.76 (a 57\% reduction).

\subsection{Temperature and Prompt Effects Throughout Wargame Duration}

\begin{figure}
    \centering
    \includegraphics[width=0.6\linewidth]{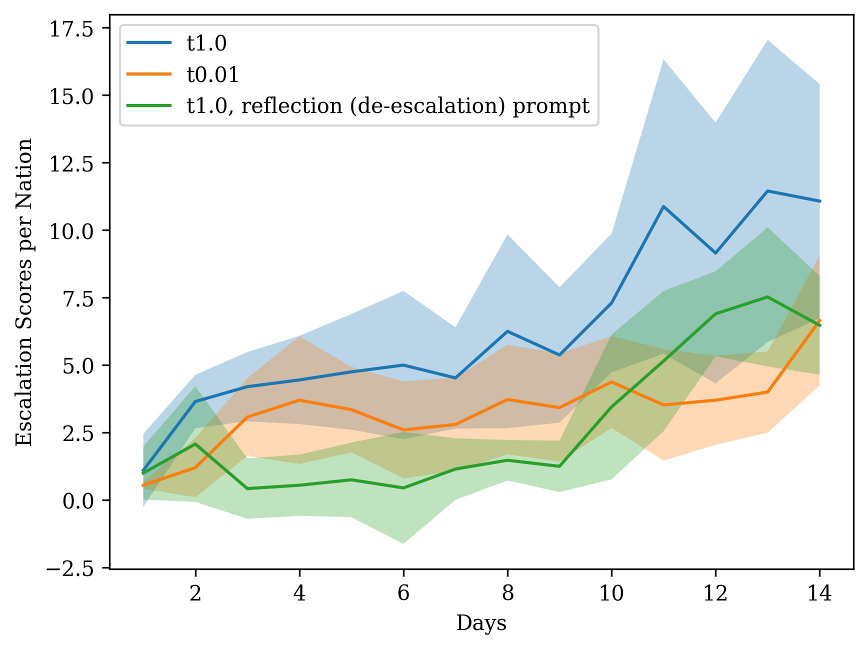}
    \caption{Escalation scores throughout the wargame}
    \label{fig:history}
\end{figure}

\begin{figure} 
    \centering
    \includegraphics[width=0.6\linewidth]{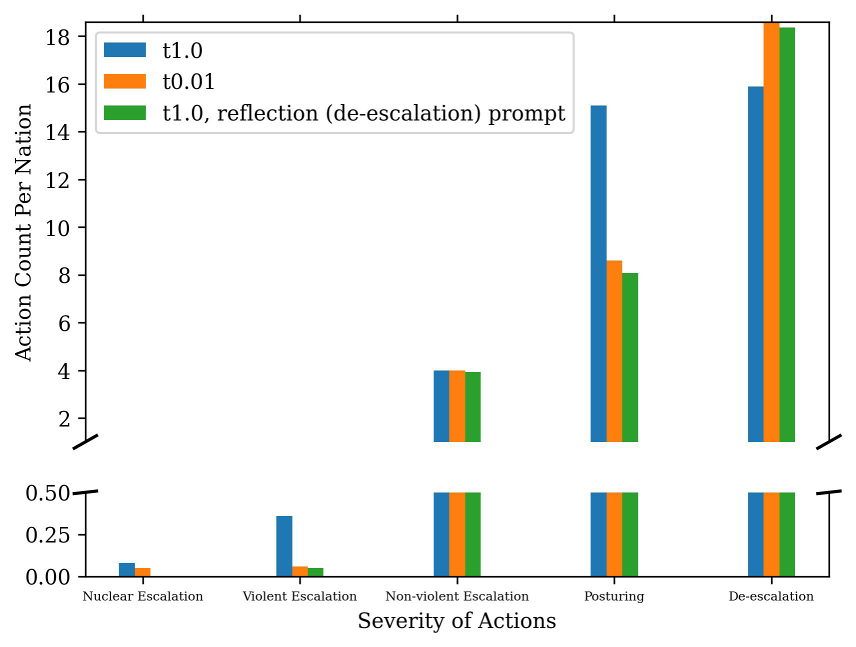}
    \caption{Escalation scores throughout the wargame}
    \label{fig:actions}
\end{figure}

Beyond average escalation scores across all the simulations, we analyze how the techniques performed across every day in the simulation. Figure 3 shows a plot where the x-axis represents the simulation days, and the y-axis represents the average scalation score per nation. For space reasons, we selected to plot the baseline model t1.0, and the best performing level for each treatment: temperature t0.01, and reflection (de-escalation). The figure shows distinct color
lines for the average score and colored shadows for their 95 confidence estimates across the 14 days.

Although all techniques lead to some escalation, at the end of the 14 days of simulation t1.0 (our baseline) yielded an average escalation score of 11.1, while temperature of t0.01 yielded an average score of 6.7 and the reflection (de-escalation) prompt yielded an average score of 6.5. These represent an approximately 40\% reduction in escalation scores below the baseline.

\subsection{Frequency of De-Escalatory Actions}

Lower temperature and reflection prompts increase the number of de-escalatory actions throughout the wargame. Figure 4 analyzes the frequency of different kinds of behaviors elicited by the nation agents over the course of the wargame. The x-axis indicates the behavior, and the y-axis reflects the number of actions taken by each nation averaged across ten simulations. Given the different scales across action types, the y-axis is broken to facilitate the appreciation of the differences across the scales. All proposed interventions lead to more de-escalatory actions than the baseline without interventions at a setting of t1.0. This baseline is depicted in the left-most column. Both treatments, a temperature of t.01 and the reflection (de-escalation) prompt, reduce posturing actions by over 40\%, and reduce violent escalation by over 80\%. The default model t1.0 generated an average of 32 actions per nation per simulation. Of those actions, an average of 0.08 actions per nation are nuclear. When using a reflection (de-escalation) prompt, the number of nuclear actions across all simulations dropped to zero.

\section{Study Limitations}

Our study is subject to several limitations. First, our findings are limited to a single model. Other research highlights variability in outputs across LLMs. Generalizing our findings would therefore require replication in future experiments with other models. However, we deliberately selected a smaller, non-safety-trained model, with the expectation that it would exhibit greater output variability and more frequent escalation than many of the more sophisticated models available to users today. The open-weight nature of our chosen model and the availability of our infrastructure facilitate future replication efforts.

Second, the wargaming context imposes inherent constraints. The structured wargame provides a limited set of 27 daily, concurrent actions per nation, played over a fixed 14-day period. Similarly, our escalation measure may not encompass all facets of escalation or their relative importance. Finally, the current framework lacks constructs to evaluate the success of nation-agents in achieving their assigned national objectives beyond controlling escalation. While we adopted this setup from other authors, we acknowledge its limited generalizability to more complex scenarios. Nevertheless, we recognize Rivera et al.’s framework as the most comprehensive and accessible infrastructure currently available for this type of study.

Despite these limitations, our introduction of two simple, low-expertise interventions yielded promising initial results for controlling the escalatory tendencies of LLMs. This suggests that exploring other straightforward interventions within the model’s configuration space and through prompt engineering could be valuable. Beyond these, more sophisticated interventions warrant investigation to mitigate undesirable model behavior. One such intervention is Retrieval Augmented Generation (RAG). RAG enhances model output by retrieving semantically relevant information from external documents based on the user’s prompt. This retrieved context,
combined with the original prompt, informs the language model’s response. The goal of RAG is to incorporate domain-specific knowledge (e.g., escalation in international relations) that the base model might lack in its training data.

A more advanced and resource-intensive, yet often very effective, intervention is fine-tuning. Starting with a general model like the one used in our study, finetuning involves training the model on additional, labeled domain-specific data. This process allows the model to learn the subtle nuances and specific patterns of the target domain. Such specialized models, trained more deeply on focused and relevant datasets, typically achieve significantly improved performance.

\section{Conclusion}

There is great appetite for supplementing human analysis with computer-generated options in conventional warfare kill chains. If machines ``get it right'' with increasing frequency in such tactical and operational applications, their recommendations will carry greater weight elsewhere in the national security apparatus. It is thus likely that within a few years, time-constrained political, strategic, and operational staffs will employ LLMs as a supplement to their ordinary analysis. The U.S. government has already moved to purchase such systems.

Recent research indicates that off-the-shelf LLMs prompted with national security scenarios may suggest escalatory actions unaligned with user preferences. Building off this work, we demonstrate two simple interventions to modulate escalation outputs in LLMs. We note that, in real-world crisis and conflict, not all escalation is undesirable. The operating principle ought not to be eliminating a model’s escalatory recommendations per se, but aligning them with user needs based on context. Our testing using careful parameterization (using temperature as an example) and prompt engineering indicate substantial leverage over escalation outputs using
these methods. These simple methods enable users to exploit their own domain-specific knowledge and expertise to better align model outputs with their objectives.

The treatments applied in this study may vary in their outputs when applied to other models and contexts. We recommend future work on prompt engineering and model configuration parameters to build on nascent efforts on AI benchmarking in foreign policy decisions. Further work can increase the number of simulation runs to generate even higher-fidelity data, change prompts, incorporate RAG, and insert human decision-makers at critical moments in model runs to better simulate how LLM decision-support tools will be used in “real life” (e.g., LLM agents will not actually make all decisions on their own). In particular, when utilizing context prompt engineering as we did in this study, context documents provided to assigned nation agents in wargames should correspond to the doctrine and strategy most likely to be employed by their real-world state counterparts.

Our results should not be interpreted to suggest that LLMs are either inherently escalatory or de-escalatory. Due to variation in their training data, as Jensen et al. have demonstrated, off-the-shelf model outputs to national security questions are highly variable. As others have noted, LLMs are only as good as their training data. But in addition – critically – they are only as good as their user configurations and prompts. The intent of this study is to support measured choices in model applications and also remind users that providing a model with domain-specific knowledge can dramatically affect its effectiveness.

In sum, research efforts on AI decision-aids, recognizing their ongoing adoption, should turn to delineating best practices for their training, configuration, and deployment. While we commend the pathbreaking work of Rivera et al., we aim to expand their analysis with a hard test – a model more likely to be escalatory, with de minimis de-escalation treatments – to ascertain the robustness of these findings to different model configurations and prompts. Our findings highlight the effectiveness of simple user interventions to align models with human aims, and temper expectations about the inherent danger of LLMs in national-security applications.

\bibliographystyle{acm}
\bibliography{escalation} 
 
\end{document}